# Development of Learning Analytics-Moodle Extension for Easy JavaScript Simulation (EjsS) Virtual Laboratories

Felix J. Garcia Clemente

*Department of Computer Engineering and Technology, University of Murcia, Murcia, Spain*

Loo Kang Wee

*Educational Technology Division, Ministry of Education, Singapore*

Francisco Esquembre

*Department of Mathematics, University of Murcia, Murcia, Spain*

Tze Kwang Leong, Darren Tan

*Curriculum Planning and Development Division, Ministry of Education, Singapore*

**Abstract**. Easy JavaScript Simulations (EjsS) is a popular and powerful authoring toolkit for the creation of open source HTML5-compliant JavaScript simulations. This paper focuses on developing a Learning Analytics extension in Moodle for EjsS, capable of monitoring interactions with the simulation (e.g. mouse clicks, states of buttons and sliders, variable assignments). This extension was piloted with educational physics simulations. Data on learners can be visualised in real-time on the instructor dashboard, allowing instructors to better understand the learning process and modify classroom instruction accordingly.

## 1 Introduction

Easy JavaScript Simulations (EjsS) [1] is an authoring toolkit that provides instructors with a click-and-drag interface for generating HTML5 JavaScript simulations or virtual laboratories. These simulations are open-source, easily edited and hosted on Open Source Physics collections at the American Association of Physics Teachers AAPT-ComPADRE[1] and the Singapore digital library[2]. Online simulations are ideally suited to capture and harness the wealth of digital data on the learning process, in what is known as learning analytics[3] – the data-driven analysis of learning activities and environments. In our work, we prototyped the Learning Analytics-Moodle extension for EjsS, which is capable of monitoring learning activities (interactions such as mouse clicks, states of the combo boxes, sliders, input to variables, number of attempts, etc) on our educational simulations in a Moodle course.

## 2 Learning Analytics-Moodle Extension for EjsS Virtual Laboratories

Incorporating learning analytics into the EjsS authoring toolkit as a Moodle extension allows student data to be captured, so that instructors can make evidence-based teaching decisions, both in real-time as well as for future lessons. For example, instructors might have an idea of what the ideal use of simulations might look like, and classify students based on whether they display such patterns of interaction with the simulations.

In addition, for distance-learning scenarios where obtaining such feedback is even more valuable, this learning analytics extension could help inform and improve simulation design.

---

[1] https://www.compadre.org/osp/
[2] https://iwant2study.org/ospsg/
[3] http://www.open.ac.uk/blogs/innovating/?page_id=582

Analysing the detailed log of all user interactions with the simulation could provide insight into possible learning challenges or issues with navigating the simulation.

The Moodle extension that we have developed monitors and records, in real-time, all the student interactions with the Easy JavaScript Simulation. This information is displayed on the instructor's dashboard with a metric of each individual students' degree of being "on task" and also shows students' current state on the simulation.

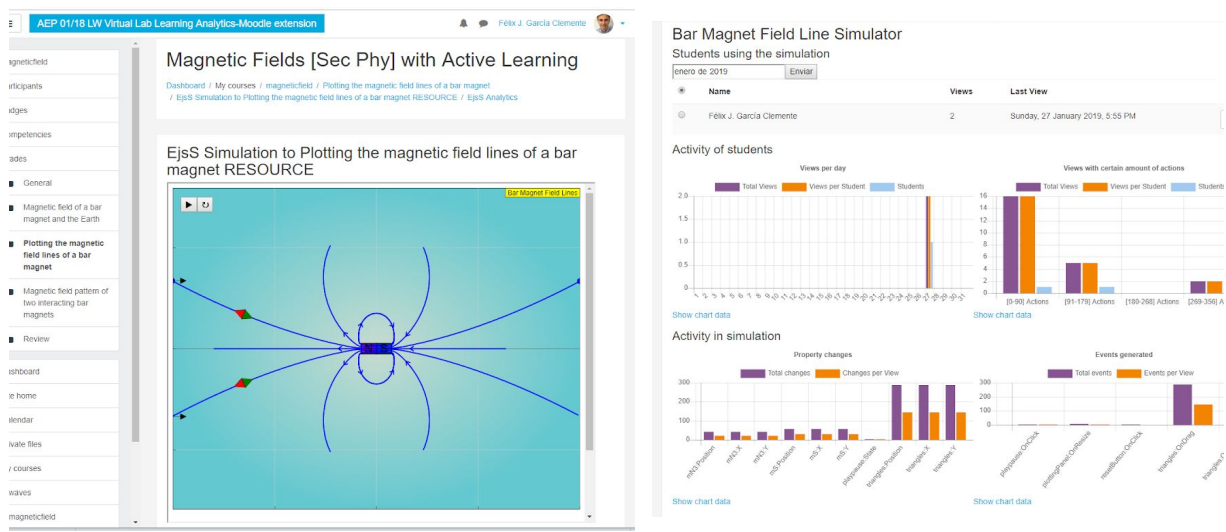

Fig. 1 EjsS Simulation (left) and the Student Interaction Summary displayed in Moodle (right)

The Moodle extension (plugin) is installed on the Open Source Physics at Singapore (OSP@SG) server[4], and provides learning analytics capability for teachers running courses based on Easy JavaScript Simulation (EjsS). This data-rich Virtual Lab functionality can be extended to existing EjsS simulations by regenerating the simulation from the source code with version 6.0 of the EjsS authoring toolkit, installing the Moodle plugin and using the simulation in a Moodle course.

## 3  Conclusion

This paper demonstrates the potential affordance of including a learning analytics extension to EjsS simulations, which can be enabled in version 6.0 of the EjsS authoring toolkit. This Moodle[5] extension we hope can provide a valuable source of useful data for instructors and learning designers to make decisions in the classroom and beyond.

**Acknowledgments**

---

[4] https://iwant2study.org/moodle
[5] https://github.com/felixgarcia/ejss-moodle-plugin